\documentstyle[prl,multicol,epsf,epsfig,aps]{revtex}
\newcommand{\BEQ}{\begin{equation}}
\newcommand{\EEQ}{\end{equation}}
\newcommand{\BEA}{\begin{eqnarray}}
\newcommand{\EEA}{\end{eqnarray}}

\renewcommand{\d}{{\rm d}}

\begin{document}
\title{Reply on Comment on
``Extraction of work from
a single thermal bath in the quantum regime''}
\date{\today}

\maketitle

\begin{abstract}
In our recent letter \cite{AN} we discussed that thermodynamics 
is violated in quantum Brownian motion beyond the weak
coupling limit. In his comment, Tasaki \cite{T}
derives an inequality for the relative entropy and
claims, without making any dynamical assumption,  
that the Clausius inequality is valid, thus contradicting our
statements\cite{AN}. 
Here we point out that the claim is unfunded, since
the author did not properly identify the
concept of heat. Tasaki also applies the inequality 
to Thomson's formulation of the second law.
This application is invalid as well, since the author did 
not correctly identify the concept of work.
Therefore, Tasaki's inequality is 
perfectly compatible with our findings.
\end{abstract}

\begin{multicols}{2}

To facilitate the reading of the present reply, we will 
use some notations of Tasaki.
As in \cite{AN,T}, we consider a quantum system
which consists of a subsystem and a bath.
The Hamiltonian of the subsystem
$H_{\rm s}(t) $
depends on time through some of its parameters;
that of the bath $H_{\rm b} $
is time-independent, and that
for interaction $H_{\rm int}(t)$
may be time-dependent.
The total Hamiltonian is
$ H(t)=H_{\rm s}(t)\otimes{\bf 1}+
{\bf 1}\otimes H_{\rm b} +
H_{\rm int}(t)$.
Initially the subsystem is in an arbitrary
equilibrium state with density matrix
$ \rho^{\rm init}_{\rm s} $,
and the bath is in the Gibbs state with
inverse temperature $\beta $.
The density matrix for the whole system is
$\rho^{\rm init}
	=
	\rho^{\rm init}_{\rm s}
	\otimes
	\exp[-\beta H_{\rm b}]/Z(\beta)$,
where \( Z(\beta)={\rm Tr}_{\rm b}
[\exp\{-\beta H_{\rm b}\}] \)
is the partition function for the bath.
(\( \rm Tr_{\rm s} \),
\( \rm Tr_{\rm b} \) stand for the traces over the spaces
of the system and the bath, and
\( \rm Tr \) indicates the full trace.)
Let \( \rho^{\rm fin} \) be the
density matrix at the final moment
obtained from the time evolution
according to \( H(t) \).
Tasaki makes {\em no assumptions} about the
nature of the time evolution or of the
final state \( \rho^{\rm fin} \).
Using well-known properties of the quantum relative 
entropy, he then derives the following inequality
\begin{equation}
	S_{\rm vN}[\rho^{\rm fin}_{\rm s}]
	-
	S_{\rm vN}[\rho^{\rm init}_{\rm s}]
	\ge
	\beta\left[
	\langle H_{\rm b}\rangle_{\rm init}
	-
	\langle H_{\rm b}\rangle_{\rm fin}
	\right],
	\label{clausius}
\end{equation}
where
\( S_{\rm vN}[\rho_{\rm s}]
=-{\rm Tr}_{\rm s}
[\rho_{\rm s}\log\rho_{\rm s}] \)
is the von Neuman entropy of the subsystem.
Then Tasaki claims that
\( \langle H_{\rm b}\rangle_{\rm init}
-\langle H_{\rm b}\rangle_{\rm fin} \)
is the {\em heat}
that flowed out of the bath
during the process. Therefore, for him, Eq.~(\ref{clausius})
becomes the Clausius inequality.

We have no doubts on the mathematical correctness of Eq.~(\ref{clausius}).
Nevertheless we colpletely disagree with identification of 
\( \langle H_{\rm b}\rangle_{\rm init}
-\langle H_{\rm b}\rangle_{\rm fin} \) as heat. 
Let first indicate that in \cite{AN} we have used a completely different
definition, which appears to be widely accepted in literature \cite{klim}. 
We will show that this common definition never coincides with that 
given by Tasaki. Then we will argue that the definition of Tasaki 
is hardly physical.

Since the complete system is closed, its dynamics is given by the 
von Neumann equation
\begin{equation}
	\frac{\d \rho (t)}{\d t}=-\frac{i}{\hbar}[H(t)\rho(t)-\rho(t)H(t)]
\label{akan}
\end{equation}
It is easy to see from Eq.~(\ref{akan}) that 
${\rm Tr}(H\frac{\d}{\d t}\rho )\equiv\d Q_{\rm tot}=0$, 
which then implies
\BEA
\label{ku}
&&0=Q_{\rm tot}=\int _{t^{\rm in}}^t\d Q_{\rm tot}=
-\Delta Q_0+\Delta Q_{\rm s}+\Delta Q_{\rm int},\\
&&\Delta Q_0=\langle H_{\rm b}\rangle _{\rm init}
-\langle H_{\rm b}\rangle_{\rm fin},\\
&&\Delta Q_{\rm s}=\int _{t^{\rm in}}^t\d t'~ {\rm Tr}_{\rm s }
[H_{\rm s}(t')\frac{\d}{\d t'}\rho _{\rm s}(t')],
\\
&&\Delta Q_{\rm int}=\int _{t^{\rm in}}^t\d t'~ {\rm Tr}
[H_{\rm int}(t')\frac{\d}{\d t'}\rho (t')]
\EEA
The interpretation of these equations is straightforward.
$\Delta Q_{\rm tot}$ is the heat obtained by the total system, and it is
zero
since the system is closed. $\Delta Q_{\rm s}$ is heat gotten by the 
particle from the thermal bath, since if there is no interaction with
the bath, the evolution of the subsystem is by itself unitary, and 
this quantity is zero. $\Delta Q_{\rm s}$ directly appears in the first
law as the energy of the subsystem \cite{AN,klim}. Moreover, it contains
only characteristics of the susbsytem, the observable ones.
Therefore, it can be controlled and measured in experiment. Taking
into account all these factors, as well as experimental confirmations,
a great amount of scientists got a conclusion that this quantity has to
be identified with heat \cite{klim}.

$\Delta Q_0$ is the quantity proposed as heat by Tasaki.
This will be an alternative definition if one can show that there
are sensible limits, where $\Delta Q_0=\Delta Q_{\rm s }$. 
However, it is easy to see that this is not the case, 
since $\Delta Q_{\rm int}$ is never zero, except when $H_{\rm int}
\to 0$, where all terms in r.h.s. of Eq.~(\ref{ku}) are zero separately,
and there is no reason to speak of heat. The situation will
not change if $H_{\rm int}$ will be put to zero starting from some control
time, since $\Delta Q_{\rm int}$ is still non-zero. 
On the other hand, the definition of Tasaki involves degrees of freedom
of the bath, which are unobservable and uncontrollable by the very
definition 
of the problem. So $\Delta Q_0$ can even not be observed directly.
It also does not appear in the first law for the particle, whereas the
standard definition does \cite{AN,klim}.

There is nevertheless some small space for a 
{\it modification} of Tasaki's definition. It arises
when one is interested in a small amount of heat obtained between
$t$ and $t+\d t$, where $t\gg t^{\rm in}$. Having taken the standard
weak-coupling assumption \cite{gardiner}, 
one considers $H_{\rm int}$ as small, and
then {\it approximately} puts 
$\d Q_0={\rm Tr}_{\rm b}[H_{\rm b}\d \rho _{\rm b}]$
equal to $\d Q_{\rm s}={\rm Tr}_{\rm s}[H_{\rm s}\d \rho _{\rm s}]$.
In this weak-coupling limit
we found that the Clausius inequality is valid \cite{AN}.
The result of Tasaki does not bear on this case, since
for his derivation it seems to be important to start at $t=t^{\rm in}$ and
consider the integral amount of heat. But the weak coupling assumption
does not apply at early times.
Thus his result is irrelevant even for this standard limit, where
the Clausius inequality is known to be valid. The reason is that
the author did not take into account any dynamical factor.
We will be surprised if Tasaki would be able to derive the proper 
Clausius inequality without making any dynamical
assumptions, since all derivations 
known to us use certain, although not the same,
set of assumptions (see \cite{A} and refs. there). 
In this context we recall that our derivations
in \cite{AN} are exact (although model-dependent), and concern the
case, where those assumptions are invalid.

Let us now mention that Tasaki uses Eq.~(\ref{clausius}) to derive
the impossibility of the perpetuum mobile, which he understands
as a violation of the Thomson's formulation of the second law:
{\it No work can be extracted from the thermal bath and subsystem 
during a cyclical process},
by a large number of identical circles. 
In this context $\Delta Q_0$ appears for him also
as the extracted work. This is again incorrect, since 
if a parameter $\alpha $ of the subsystem is varying with time then
the extracted work $\Delta W$ is given as \cite{AN,klim}
$$\int _{t^{\rm in}}^t\d t' ~\frac{\d\alpha}{\d t'} 
{\rm Tr}_{\rm s}\left [\frac{\partial H_{\rm s}}{\partial \alpha}
\rho _{\rm s}(t')\right ]=\int _{t^{\rm in}}^t\d t' ~\frac{\d\alpha}{\d
t'} 
{\rm Tr}_{\rm s,b}\left [\frac{\partial H}{\partial \alpha}
\rho (t')\right ]
$$ 
and this is just the work extracted by external sources from the total 
system. 
Needless to mention that $\Delta Q_0\not =\Delta W$.
So this statement of Tasaki is also incorrect. 

R. Balian is acknowledged for interesting discussions.

We thank Hal Tasaki for communicating us his result before
to make it public.


{A.E. Allahverdyan$^{1,3)}$ and Th.M. Nieuwenhuizen$^{2)}$}\\

{$^{1)}$S.Ph.T., CEA Saclay, 91191 Gif-sur-Yvette cedex, France;\\

$^{2)}$Department of Physics and Astronomy,
University of Amsterdam,

Valckenierstraat 65, 1018 XE Amsterdam, The Netherlands; \\ 

$^{3)}$Yerevan Physics Institute,

Alikhanian Brothers St. 2, Yerevan 375036, Armenia. }

\end{multicols}
\end{document}